\documentclass[prl,groupedaddress,showpacs,preprintnumbers,amsmath,amssymb,twocolumn,floats]{revtex4}
\usepackage{txfonts}
\usepackage{amssymb}
\usepackage{graphicx}

\begin{document}

\title{Angle-resolved photoemission evidence of $s$-wave superconducting gap in K$_x$Fe$_{2-y}$Se$_2$ superconductor}

\author{M. Xu}
\author{Q. Q. Ge}
\author{R. Peng}
\author{Z. R. Ye} \author{Juan Jiang} \author{F. Chen} \author{X. P. Shen} \author{B. P. Xie}\author{Y. Zhang}\email{yanzhangfd@fudan.edu.cn}
\author{D. L. Feng}\email{dlfeng@fudan.edu.cn}
\affiliation{State Key Laboratory of Surface Physics, Department of Physics,  and Advanced Materials Laboratory, Fudan
University, Shanghai 200433, People's Republic of China}

\begin{abstract}
Although nodeless superconducting gap  has been observed on the large Fermi pockets around the zone corner in K$_x$Fe$_{2-y}$Se$_2$, whether its pairing symmetry is $s$-wave or nodeless $d$-wave is still under intense debate. Here we report an isotropic superconducting gap distribution on the small electron Fermi pocket around the Z point in
K$_x$Fe$_{2-y}$Se$_2$, which favors the $s$-wave pairing symmetry.
\end{abstract}

\maketitle

Compared with the cuprate superconductors, the iron-based  high temperature superconductors (Fe-HTS's) exhibit much more diversified structures and electronic structures, which provides a rich playground for various competing  instabilities. As a result, the  pairing symmetry of the Cooper pair, one of the most fundamental characters of a superconductor,  is still far from well understood for Fe-HTS's. For the Fe-HTS's whose Fermi surface is composed of hole pockets in the Brillouin zone center and electron pockets at the zone corner,
$s$-wave pairing has been generally established not only in systems with nodelss superconducting gap, such as  Ba$_{1-x}$K$_x$Fe$_2$As$_2$,  BaFe$_{2-x}$Co$_{x}$As$_2$,  FeTe$_{1-x}$Se$_x$  \cite{nodeless1,nodeless2,DingFeSe,ZhangBK,HanaguriSTM}, but also even in ones with nodal superconducting gap, such as  BaFe$_2$(As$_{1-x}$P$_x$)$_2$ \cite{ZhangNatPhys},  and heavily overdoped  Ba$_{1-x}$K$_x$Fe$_2$As$_2$ \cite{Shinnew}. However, theoretically, whether there is a sign change or not for the order parameter in different Fermi sheets is still unsettled. For example, it could depend on whether spin fluctuations \cite{Kuroki,Mazin} or orbital fluctuations \cite{Kontani} are taken as the dominating pairing force. Moreover, for the recently discovered A$_x$Fe$_{2-y}$Se$_2$ (A= K, Cs, Rb, ...) superconductor \cite{ChenXL,FangMH1,ChenXH2}, its parent compound \cite{Bao1,ChfPRX} and electronic structure  are both rather different from other Fe-HTS's \cite{ZhangNM,ZhouSe122,DingSe122}, and whether its pairing symmetry is $s$-wave  or nodeless $d$-wave is currently under intense debate \cite{Hu,Hu0,Hu1,Hu2,Hu3,Swave1,dwave1,dwave2,dwave3,dwave4,dwave5}. Therefore, an unified theme for the pairing symmetry in the Fe-HTS's have not been reached.

\begin{figure}[t]
\includegraphics[width=8.6cm]{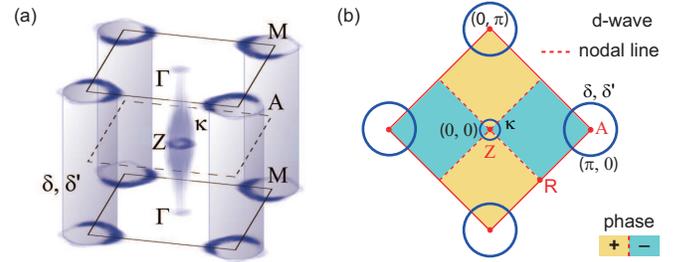}
\caption{(a) Schematic diagram of the Fermi surface of K$_x$Fe$_{2-y}$Se$_2$ superconductor in the 3D Brillouin zone corresponding to two iron ions per unit cell.  (b) Schematic diagram of the $d$-wave theoretically gap symmetry with the nodal lines along the $(0,0)-(\pm\pi,\pm\pi)$ directions. Positive and negative phases of the superconducting order parameter is denoted by different colors. } \label{intro}
\end{figure}

Taking K$_x$Fe$_{2-y}$Se$_2$ as an example, its Fermi surface is made of a small electron pocket ($\kappa$) around Z, and  large electron cylinders ($\delta$ and $\delta'$) at the zone corners, as reported earlier and shown in Fig.~\ref{intro}(a) \cite{ChfPRX}.  A large and isotropic superconducting gap of $\sim 10 meV$ has been observed for the $\delta$/$\delta'$ electron pockets, while a gap of $ \sim 8$~meV has been observed on the $\kappa$ pocket \cite{ZhangNM}. The lack of hole pockets near the zone center rules out the common s$^{\pm}$ pairing symmetry proposal based on scattering between the hole and electron Fermi surface sheets in the iron-pnictides. Theories based on local antiferromagnetic exchange interactions have predicted $s$-wave pairing symmetry in this system that can account for the experiments \cite{Hu,Hu0,Hu1,Hu2,Hu3,Swave1}. However, calculations based on the scattering amongst the $\delta$/$\delta'$ electron pockets have indicated that the $d$-wave pairing channel would win over the $s$-wave pairing channel \cite{dwave1,dwave2,dwave3,dwave4,dwave5}, which makes the superconducting order parameters to change sign between neighboring  $\delta$/$\delta'$ Fermi pockets as illustrated in Fig.~\ref{intro}(b). Because the nodes would appear along the four $(0,0)-(\pm\pi,\pm\pi)$ directions (dashed lines) that do not cross any of the $\delta$/$\delta'$ Fermi cylinders,  it can explain the observed nodeless gap structure on the $\delta$/$\delta'$ electron pockets.  So far, these two pairing scenarios have not been distinguished experimentally.

Noting that the nodal lines in the $d$-wave pairing scenario actually cross the $\kappa$  pocket, one would expect nodes in the superconducting gap at such crossings, whereas one would expect the gap to be nodeless in the $s$-wave  pairing scenario. Therefore, the superconducting gap distribution along the $\kappa$ pocket could serve as a benchmark to determine the pairing symmetry of K$_x$Fe$_{2-y}$Se$_2$. Previously, because of the small size of the $\kappa$ Fermi pocket, the detailed distribution of its gap has not been reported. We here report the angle-resolved photoemission spectroscopy (ARPES)
measurements on the detailed superconducting gap on the $\kappa$ Fermi pocket in
K$_x$Fe$_{2-y}$Se$_2$ superconductor. We found an isotropic superconducting gap of $\sim8$meV, which favors the $s$-wave pairing symmetry. This work together with previous studies suggest that the pairing symmetry of all the Fe-HTS's discovered so far is ubiquitously $s$-wave.

\begin{figure}[t]
\includegraphics[width=8.6cm]{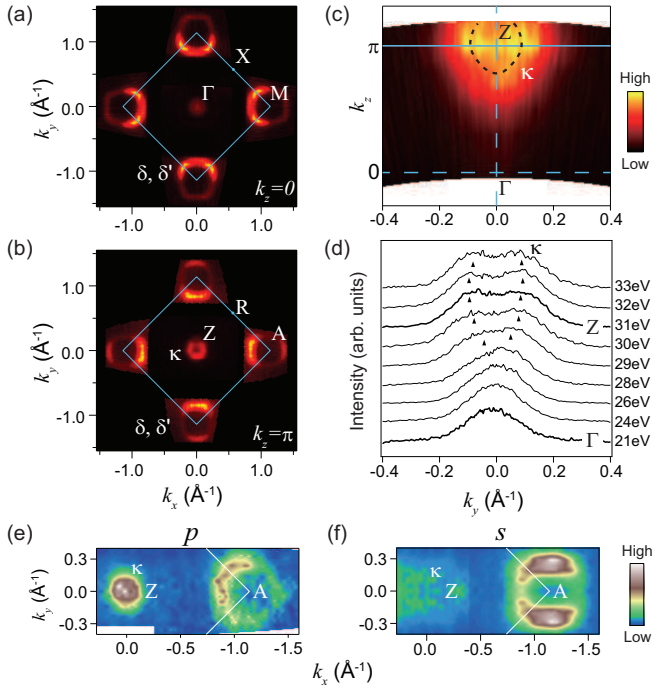}
\caption{Fermi surface of  K$_x$Fe$_{2-y}$Se$_2$ superconductor. (a) The photoemission intensity map for the Fermi surface at $k_z$~=~0 taken with 21~eV photons at 35~K.  The $k_x$ and $k_y$ are defined along Fe-Fe directions. (b) The photoemission intensity map for the Fermi surface at $k_z$~=~$\pi$ taken with 31eV photons at 35~K. (c) The photoemission intensity map taken along $k_y$-$k_z$ cross-section of the three-dimensional (3D) Brillouin zone. (d) The momentum distribution curves (MDCs) at Fermi energy ($E_F$) taken at different photon energies. (e), (f) The polarization dependence of the photoemission intensity maps taken in the Z-A plane at 35K taken with the $p$ and $s$ polarization geometries respectively. Data were taken at SSRL for panels a, b, c, and d, and UVSOR for panels e and f.} \label{fs}
\end{figure}

K$_x$Fe$_{2-y}$Se$_2$ single crystals were synthesized by self-flux method as described elsewhere in detail \cite{ChenXH2}, which show flat shiny surfaces with dark black color. The superconducting sample shows the superconducting transition temperature of $\sim$31~K.   The  chemical compositions of the samples  were determined by energy dispersive X-ray (EDX) spectroscopy to be  K$_{0.77}$Fe$_{1.65}$Se$_2$. The synchrotron ARPES experiments were performed at Beamline 5-4 of SSRL synchrotron facility, and Beamline 21B1 of NSRRC facility, with Scienta R4000 electron analyzers. The overall energy resolution in the gap measurement is about 10~meV at SSRL or NSRRC, and angular resolution is 0.3 degree. The polarization dependence of the Fermi surface  was measured at UVSOR facility with an MBS A-1 electron analyzer. The samples were cleaved \textit{in situ}, and measured under ultra-high-vacuum of $5\times10^{-11}$\textit{torr}.

\begin{figure}[t]
\includegraphics[width=8.6cm]{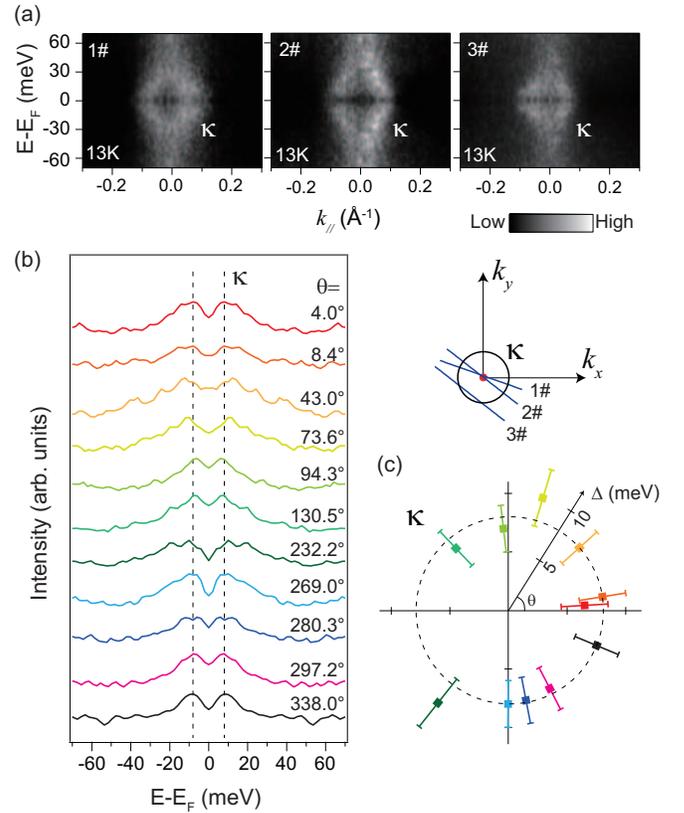}
\caption{The superconducting gap at the $\kappa$ pocket of K$_x$Fe$_{2-y}$Se$_2$.
(a) Symmetrized photoemission intensities for three momentum cuts across the $\kappa$ pocket near Z as shown by the thick lines \#1-\#3 in the inset. (b)
Symmetrized energy distribution curves at the Fermi crossings of the $\kappa$ band with momenta counterclockwise along the $\kappa$ pocket as shown by the labeled polar angles. (c) Gap distribution of the $\kappa$ pocket around Z in polar coordinates, where the radius represents the gap size, and the polar angle $\theta$ represents the position on the $\kappa$ pocket with respect to the Z point, $\theta$=0 being the $\Gamma$-M direction.  The data were taken at NSRRC with 31~eV photons at 13~K.} \label{gap}
\end{figure}

For a close examination of the $\kappa$ electron pocket, Figs.~\ref{fs}(a)-\ref{fs}(b) show the photoemission intensity maps for the Fermi surface at two different $k_z$'s. While the electron pockets around the zone corner show little $k_z$ dependence, the small $\kappa$ pocket could be only observed near the Z point for $k_z$~=~$\pi$, indicating its strong three-dimensional (3D) character. The $k_z$ dependence of $\kappa$ is further illustrated by the photoemission intensity map along the $k_y$-$k_z$ cross-section of the Brillouin zone [Fig.~\ref{fs}(c)]. We determined the Fermi crossings of $\kappa$ according to the momentum distribution curves (MDCs) near Fermi energy ($E_F$) taken with different photon energies [Fig.~\ref{fs}(d)]. Such a small enclosed $\kappa$ electron pocket around Z is generally overlooked by theories in studying the pairing symmetry of K$_x$Fe$_{2-y}$Se$_2$, even it was found to exhibit a superconducting gap with comparable amplitude as that on the large $\delta$/$\delta'$ electron cylinders at the zone corner.

Previous photoemission studies on  K$_x$Fe$_{2-y}$Se$_2$ only resolved one electron pocket around the zone corner. In the folded Brillouin zone with two iron ions per unit cell, there should be two electron cylinders around the zone corner due to folding. Band calculations show that these two electron cylinders in  K$_x$Fe$_{2-y}$Se$_2$  have opposite symmetries with respect to the $Z-\Gamma-M$ plane \cite{bandKFe2Se2}, as found in other Fe-HTS's. To resolve this discrepancy, we show the polarization-dependent ARPES data in Figs.~\ref{fs}(e) and \ref{fs}(f). Due to the multi-orbital nature of the Fermi surface in iron-based superconductors, certain Fermi surface sheet might exhibit either even or odd spatial symmetry, thus could be observed in the $p$ and $s$ polarization geometries respectively \cite{zhangbaco}. In the polarization-dependent ARPES studies of BaFe$_{2-x}$Co$_{x}$As$_2$ and NaFeAs, one square-like electron pocket and one elliptical electron pocket could be separately probed around the zone corner in the $p$ and $s$ polarization geometries, respectively \cite{zhangbaco,ZhangNaFeAs}. Here in Figs.~\ref{fs}(e) and \ref{fs}(f), there is an electron pocket around A in both the $p$ and $s$ polarization geometries, indicating the existence of two electron pockets with opposite symmetries. Moreover, unlike the electron pockets observed in other Fe-HTS's, the two electron pockets could not be resolved in momentum, which indicates a low-ellipticity and high degeneracy of these two electron pockets in K$_x$Fe$_{2-y}$Se$_2$. This  explains why only one electron pocket could be observed with the mixed polarization geometry in previous photoemission studies.

In the $d$-wave pairing scenario, the superconducting order parameter in the neighboring quarters has opposite signs in the unfolded Brillouin zone. It has been argued that in the folded Brillouin zone, where there are two iron ions in one unit cell, the folding of electron cylinders from one quarter to its neighboring quarter could induce hybridization and thus extra nodes on themselves \cite{Mazinphysics}. However, this might not be the case here, due to the low-ellipticity and weak interactions between the $\delta$/$\delta'$ electron pockets. Therefore, the isotropic and nodeless superconducting gap distributions on the $\delta$/$\delta'$ pockets might not be viewed as a conclusive evidence against the $d$-wave pairing symmetry in K$_x$Fe$_{2-y}$Se$_2$.

\begin{figure}[t]
\includegraphics[width=6.6cm]{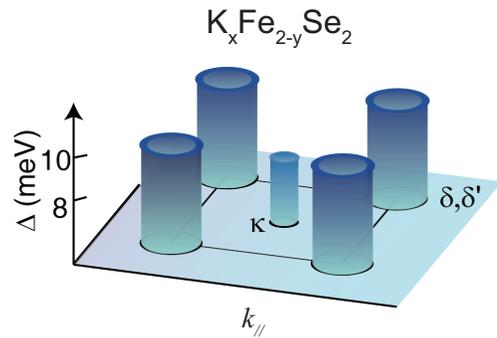}
\caption{ Summary of the gap distribution at $\kappa$, $\delta$, and $\delta'$ pockets in K$_x$Fe$_{2-y}$Se$_2$.} \label{sum}
\end{figure}

We now turn to the detailed distribution of the superconducting gap around the $\kappa$ pocket. Fig.~\ref{gap}(a) shows the symmetrized photoemission intensities along several cuts across the $\kappa$ pocket near Z. The electron-like band dispersion is clearly visible, and the dip in intensity at the $E_F$ illustrates the opening of a superconducting gap. Fig.~\ref{gap}(b) plots the spectra in the
superconducting state at various Fermi crossings of $\kappa$ around
the Z point. Clear superconducting coherence peaks were observed at
different momenta and they roughly share a same peak position of about
$\sim8$~meV as marked by the vertical dashed line in Fig.~\ref{gap}(b). The
momentum distribution of the superconducting gap on $\kappa$ is shown in
Fig.~\ref{gap}(c), which clearly shows a nodeless isotropic superconducting
gap distribution. Note that the momenta we chose here almost cover
the entire $\kappa$ pocket, so the gap nodes should not be missed if they had existed. Together with the nodeless superconducting gap on $\delta$/
$\delta$' electron pockets determined by previous ARPES studies \cite{ZhangNM}, we
conclude a nodeless superconducting gap structure in all Fermi surface sheets of
K$_x$Fe$_{2-y}$Se$_2$ as summarized in Fig.~\ref{sum}.


The observed nodeless isotropic superconducting gap on the $\kappa$ Fermi pocket is consistent with the $s$-wave pairing symmetry, while it poses serious challenges to the  $d$-wave pairing scenario  in K$_x$Fe$_{2-y}$Se$_2$. Although the existing theories that predicted $d$-wave have neglected the contribution of the $\kappa$ pocket, the interactions between  $\delta$/$\delta'$  and  $\kappa$ would most likely induce a phase change and thus nodes on $\kappa$. Considering the small density of states of $\kappa$, its large gap size is likely caused by such a proximity effect from the  $\delta$/$\delta'$. It is also probable that there is a  $d$-wave pairing component in addition to the $s$-wave one, for example, in the $s+id$ pairing form, however, the isotropic superconducting gaps on all the Fermi surface sheets indicate that the $d$ component should be very weak.

To summarize, we have identified isotropic nodeless superconducting gap distribution on the $\kappa$ Fermi pocket in K$_x$Fe$_{2-y}$Se$_2$, which provides a strong experimental support to the $s$-wave pairing symmetry in this compound. Our results indicate $s$-wave is likely a ubiquitous pairing symmetry of iron-based high temperature superconductors in general.

We are grateful to Dr. Donghui Lu for experimental assistance at SSRL, and Prof. Jiangping Hu, Prof. Zidan Wang, Prof. J. X. Li, and Prof. Q. H. Wang for helpful discussions. This work is supported in part by the National Science Foundation of China, and National Basic Research Program of China (973 Program)  under the grant  Nos. 2012CB921400, 2011CB921802 and 2011CBA00112. SSRL is operated by the US DOE, Office of Basic Energy Science, Divisions of Chemical Sciences and Material Sciences.

\end{document}